\documentclass[manuscript,screen]{acmart}
\AtBeginDocument{%
  \providecommand\BibTeX{{%
\normalfont B\kern-0.5em{\scshape i\kern-0.25em b}\kern-0.8em\TeX}}}
\makeatletter
\renewcommand\@formatdoi[1]{\ignorespaces}
\makeatother
\setcopyright{rightsretained}
\copyrightyear{2024}
\acmYear{2024}
\acmDOI{}
\acmConference[CCAI 2024]{CHI 2024 Workshop on Child-centred AI Design}{May 11, 2024}{Honolulu, HI, USA}
\acmBooktitle{CHI 2024 Workshop on Child-centred AI Design, May 11, 2024, Honolulu, HI, USA}
\acmISBN{}

\begin{document}

%%
%% The "title" command has an optional parameter,
%% allowing the author to define a "short title" to be used in page headers.
\title{PlayFutures: Imagining Civic Futures with AI and Puppets}

%%
%% The "author" command and its associated commands are used to define
%% the authors and their affiliations.
%% Of note is the shared affiliation of the first two authors, and the
%% "authornote" and "authornotemark" commands
%% used to denote shared contribution to the research.
\author{Supratim Pait}
\email{spait3@gatech.edu}

\affiliation{%
  \institution{Digital Media, Georgia Institute of Technology}
  \country{USA}
} 

\author{Sumita Sharma}
\affiliation{%
  \institution{INTERACT Research Unit, ITEE, University of Oulu}
  \city{Oulu}
  \country{Finland}}

\author{Ashley Frith}
\affiliation{%
  \institution{Human Computer Interaction, Georgia Institute of Technology}
  \city{Atlanta}
  \country{USA}}

\author{Michael Nitsche}
\affiliation{%
  \institution{Digital Media, Georgia Institute of Technology}
  \city{Atlanta}
  \country{USA}}

\author{Noura Howell}
\affiliation{%
  \institution{Digital Media, Georgia Institute of Technology}
  \city{Atlanta}
  \country{USA}}

\renewcommand{\shortauthors}{Pait et al.}

%%
%% By default, the full list of authors will be used in the page
%% headers. Often, this list is too long, and will overlap
%% other information printed in the page headers. This command allows
%% the author to define a more concise list
%% of authors' names for this purpose.

%%
%% The abstract is a short summary of the work to be presented in the
%% article.

\begin{abstract}
Children are the builders of the future and crucial to how the technologies around us develop. They are not voters but are a participants in how the public spaces in a city are used. Through a workshop designed around kids of age 9-12, we investigate if novel technologies like artificial intelligence can be integrated in existing ways of play and performance to 1) re-imagine the future of civic spaces, 2) reflect on these novel technologies in the process and 3) build ways of civic engagement through play. We do this using a blend AI image generation and Puppet making to ultimately build future scenarios, perform debate and discussion around the futures and reflect on AI's role and potential in their process. We present our findings of how AI helped envision these futures, aid performances, and report some initial reflections from children about the technology.
\end{abstract}

%%
%% The code below is generated by the tool at http://dl.acm.org/ccs.cfm.
%% Please copy and paste the code instead of the example below.
%%
%%
%% Keywords. The author(s) should pick words that accurately describe
%% the work being presented. Separate the keywords with commas.
%% A "teaser" image appears between the author and affiliation
%% information and the body of the document, and typically spans the
%% page.
\begin{teaserfigure}
  \includegraphics[width=\textwidth]{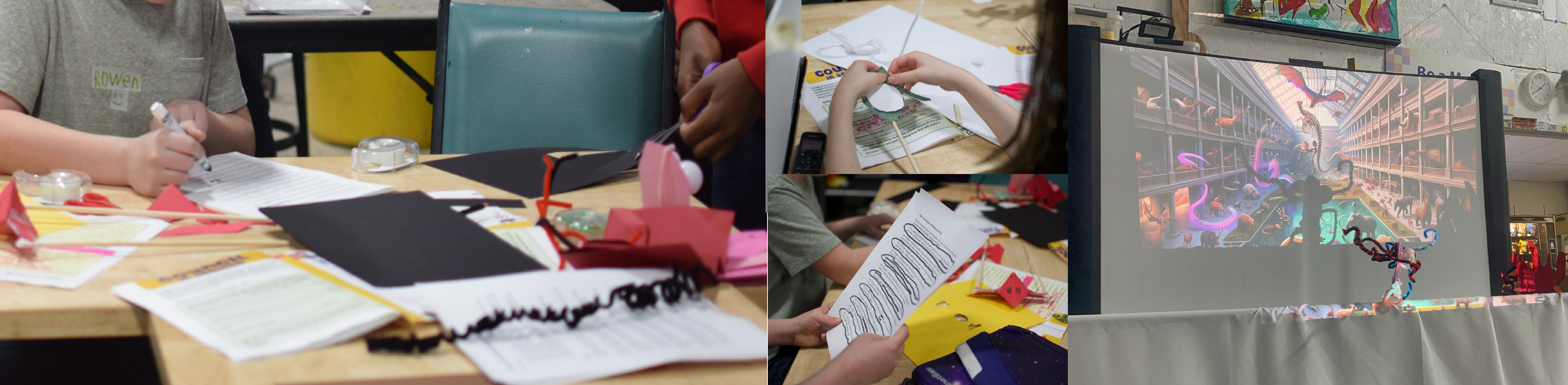}
  \caption{PlayFutures is a workshop for kids aged 10-12 yrs old on a) art making \& b) storytelling with puppets with the help of c) Generative AI Art. Children learn about about AI -Art to re-imagine civic spaces in Atlanta, GA, USA, and use AI to imagine future scenarios, make puppets and act out those stories to discuss our ideas of the future..}
  \Description{}
  \label{fig:teaser}
\end{teaserfigure}

%%
%% This command processes the author and affiliation and title
%% information and builds the first part of the formatted document.
\maketitle

\section{Introduction}

In educational settings, fostering a sense of criticality among children is increasingly recognized as essential. Iversen et al. \cite{iversen_computational_2018} point out how the importance of embedding critical thinking into technology education underscores the need to move beyond traditional computational thinking (CT) methodologies, which significantly lacks in encouraging a critical and reflective stance towards our digitized society, thereby limiting children's capacity to make informed choices about technology and participate in its development.

Participatory design (PD) emerges as a paradigm to bridge this gap, offering an avenue for children to engage critically with technology. PD, by its very nature, is inclusive and collaborative, inviting participants to actively contribute to the design process, while also bringing out aspects of their identity \cite{coenraad_enacting_2019}. This methodology aligns with the concept of computational empowerment (CE) \cite{iversen_computational_2018}, suggesting an enriched educational framework where children are not mere consumers of technology but active participants in its critique and creation. 
Design Futuring involves imagining and crafting future scenarios to inform design decisions and anticipate potential alternative realities. This practice underlines the necessity for a plurality of perspectives \cite{howell_calling_2021}, advocating for a future that is inclusively envisioned and resonates with diverse societal needs and aspirations. The importance of this pluralistic approach is echoed with the works on values in design \cite{friedman_value_2019} and reflective design \cite{sengers_reflective_2005}, emphasizing the need to challenge the status quo and envisage futures that are equitable and just.

Making becomes an essential mechanism for both exploring criticality \cite{ratto_critical_2011}and futuring. The act of making, especially within educational settings, serves as a tangible means through which children can engage with concepts of technology and futuring. It provides a hands-on experience that not only facilitates a deeper understanding of technology's potential and limitations but also empowers children to envision and shape future technological landscapes.

Performance allows children to embody and enact future scenarios, making the abstract tangible and the future immediate. \cite{schechner_performance_2013} This engagement is not merely performative but deeply participatory, fostering a connection between the children and the material, in this case, the puppets, which become proxies for exploring and critiquing future technologies and societal configurations.

Lastly, the role of civic spaces in leveraging participatory methods for engaging with technology and society is crucial. By integrating PD practices into Civic Research, particularly with children through the workshop PlayFutures, we contribute to an expanding field of work \cite{iivari_critical_2022}  that seeks to democratize technological futures and civic engagement. 

\section{Methodology}

The workshop is designed for 9-12 year old's, targeting an age where their capability for active engagement, coupled with a balance of independence and imaginative play, make them an ideal age group for critical engagement. The workshop begins with an introduction to AI image generation and public spaces, where children articulate their perceptions and aspirations for the future of communal areas.

Subsequently iterating on the images in ChatGPT, a chat based interface for the GPT-4 large language model for text and image generation, children are able to creatively add, remove, modify elements from original images of the spaces.  This is achieved through prompt structures called "spells," that guide the children in articulating their desires for visualise the change, incorporating both their individual preferences and collective decision making. 

The reflective phase prompts children to evaluate the outcomes of their image generation processes, and considering the role of AI in shaping their visions for public spaces.

The next phase, children transition from designers to storytellers and performers. They create characters who are part of the city council, with one agreeing with the re-imagined changes to the public place in the image generation phase, one who disagrees and, if they are a group of more than two, an additional character who would try to mediate the debate between the former two. The dialogue and scenes are generated ChatGPT, but with the children having the freedom to change contents.

\begin{figure}
    \centering
    \includegraphics[width=1\linewidth]{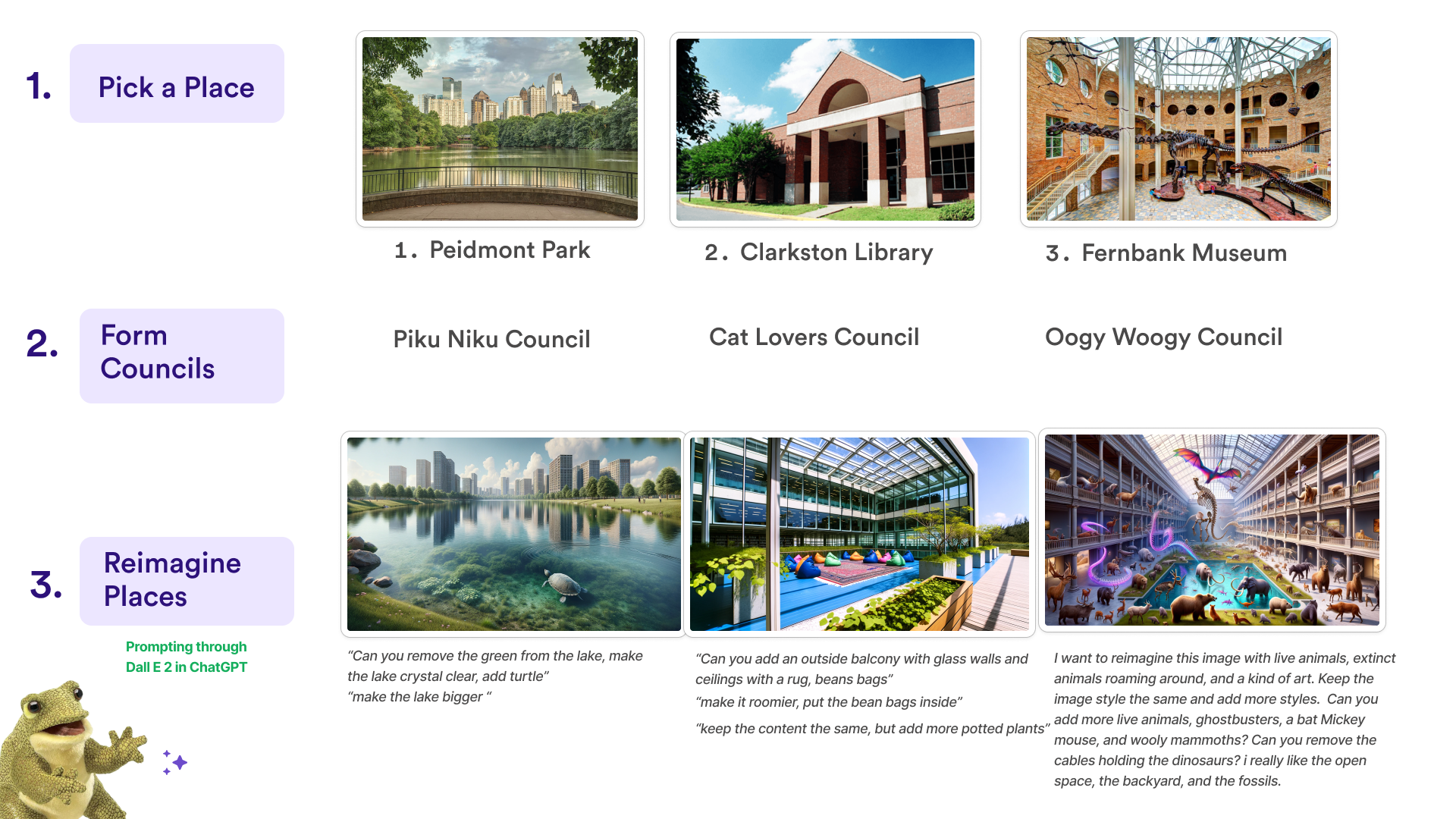}
    \caption{1) Kids discussed public places of they want to redesign and formed three council 2)\textit{The Piku Niku Council, Cat Lovers Council, }and the\textit{ Oogy Woogy Council}. 3) They used AI Image generation using prompts to re-image their chosen places throughout Atlanta, Georgia, USA. The figure also shows the kind of prompts used to approach this.}
    \label{phase-one}
\end{figure}
\begin{figure}
    \centering
    \includegraphics[width=1\linewidth]{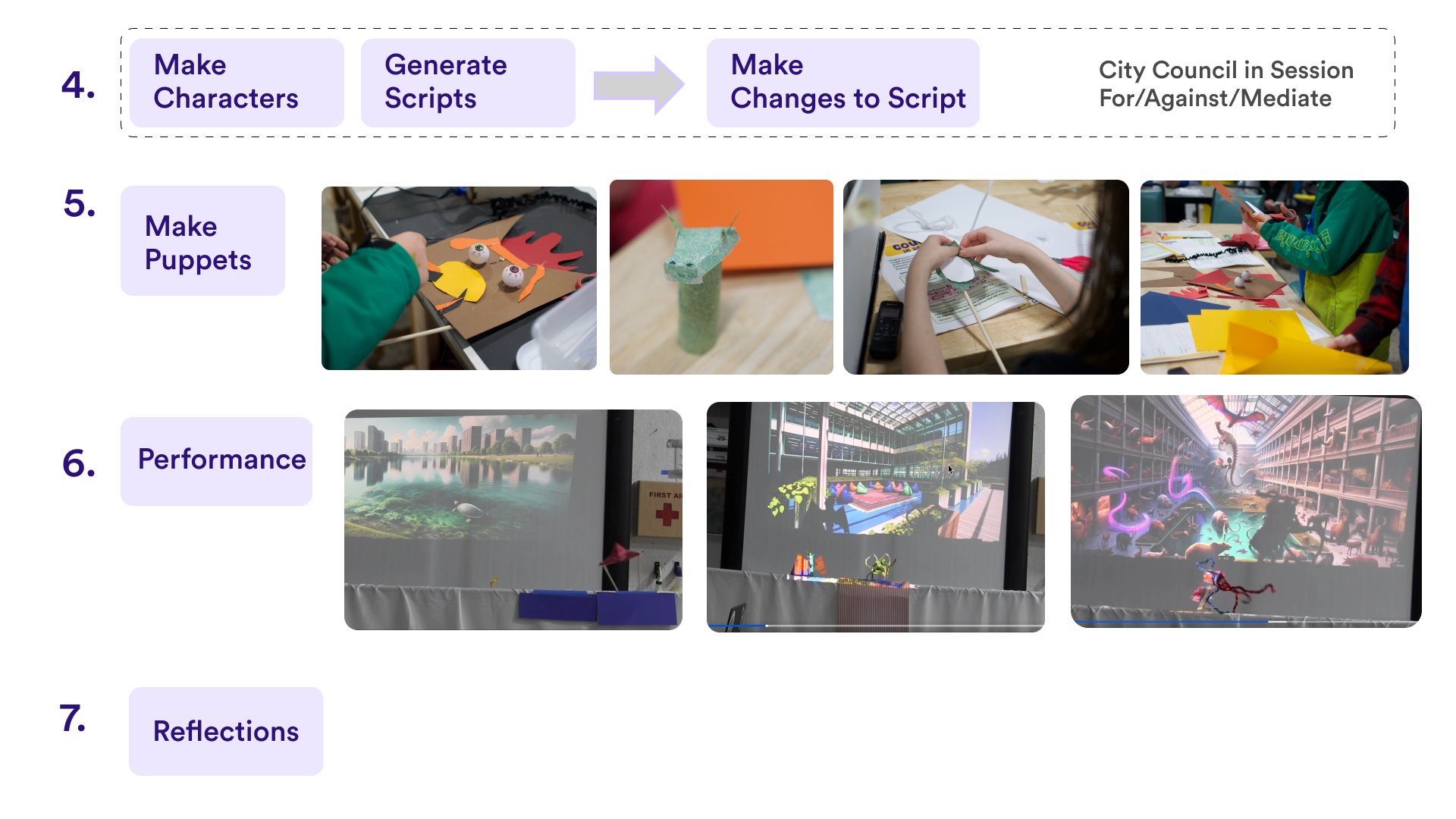}
    \caption{The second part of the workshop focused on Council in Session: where 4) participants generated city council debate scripts, starring characters they thought of, with ChatGPT. They were also told that they had the freedom to change anything about the script. 5) After an hour of making these puppets, 6) we taped the scripts with the changes behind the curtain for them to read aloud and perform }
    \label{phase-two}
\end{figure}
The workshop culminates in a play of council session, where children, through their making puppets, engage in a performance of a city council debate, enacting different perspectives of the council members on the future of the chosen space. This performance is instrumental in embodying the dialogues and tensions inherent in civic decision-making, offering insights into the complexities of participatory civic research where puppets act as a probe \cite{gaver_design_1999}. 

The workshop concludes with a reflective session, allowing participants to articulate their experiences, changes made to the AI-generated stories, and their feelings towards the futures envisioned. This final stage covers the learning outcomes and personal growth experienced by the children through the workshop.

\section{Initial Findings and Discussion}

The workshop was conducted in a community makerspace, which in turn allowed access to materials, space and equipment. An open call for children who were interested to explore Puppets and AI was made. In total, 7 Participants were present and the workshop was scheduled for around 4 hours. We formed groups, based on a common public place children were interested in, who named themselves Cat Lovers Council, Oogy Woogy Council, and Piku Niku Council. They were formed to tackle different types of public spaces, including libraries, parks, and museums, using ChatGPT for image generation and script writing. The process was hands-on and iterative, with participants engaging directly with AI to visualize their ideas and receive immediate feedback through generated images and scripts.

As seen in Fig.\ref{phase-one} the teams worked on sites of Piedmont Park, Clarkston Library and the Fernbank Museum to imagine how they envision the future of these spaces. The Cat Lovers Council focused on making a library more inviting, incorporating ideas like increased natural light and comfortable seating. Oogy Woogy group considered environmental enhancements to outdoor spaces, emphasizing sustainability. The discussion around Piedmont Park by the Piku Niku Council involved making the lake clearer, adding more life around it. Throughout these sites, the emphasis was on using AI as a tool for visualising ideas, pushing the boundaries of conventional planning and design.

Our initial findings suggest, reflecting on the image generation,  participants found value in the process beyond the immediate visualisation. Participants appreciated the excitement that came from being able to quickly visualize changes to public spaces, even if the AI's execution was not always precise. They recognized the need for clear instructions and the benefits of an iterative approach, allowing for refinement and adjustment. 

The teams learned to refine their communication with AI, generating scripts about a debate scenario in the City Council, where characters would agree and disagree with the changes. Adjustments to their scripts were based on their own preferences, allowing them the freedom to change generated content. They engaged in making the characters as puppets while making the changes along the way. These changes to the scripts become a crucial part of our reflections with them. Finally, all teams were able to perform. The image generated in the initial phases of the workshop was used as the backdrop for the performance to keep the focus on the future. At the end, all teams ended up changing elements of their scripts. One of the teams decided to improvise, completely rejecting the script, finding the script not "\textit{fun}" and  lacking "\textit{soul}". They iterated that "\textit{The whole thing felt really AI-y.}" One team agreed with the ending, while the other two completely changed it. The reflections highlighted their understanding of AI as system, with optimism about future advancements in AI tools as support.

In conclusion, the workshop served as an exploration of ways in which AI can intersect with play and performance to re-imagine public spaces. It explored play as a means to reflect and play as a way to design. Through teamwork, iterative learning, and direct engagement with AI through embodied play, participants gained insights into both the potential and challenges of leveraging AI in creative processes and the futures of civic spaces.

\section{Future Work}
This is a work in progress with preliminary findings from data.  Next steps are a more deeper analysis with more nuanced reporting of findings along with their implications to design research practices for Child Computer Interaction, Civic Research and Design Futuring.

\bibliographystyle{ACM-Reference-Format}
\bibliography{main}
\end{document}